\documentclass[a4paper,11pt]{article}
\pdfoutput=1 

\usepackage{jinstpub} 
                     
\usepackage{listings}
\usepackage{xcolor}

\lstdefinelanguage{JavaScript}{
  keywords={break, case, catch, continue, debugger, default, delete, do, else, false, finally, for, function, if, in, instanceof, new, null, return, switch, this, throw, true, try, typeof, var, void, while, with},
  morecomment=[l]{//},
  morecomment=[s]{/*}{*/},
  morestring=[b]',
  morestring=[b]",
  showstringspaces=false,
  ndkeywords={class, export, boolean, throw, implements, import, this},
  keywordstyle=\color{blue}\bfseries,
  ndkeywordstyle=\color{darkgray}\bfseries,
  identifierstyle=\color{black},
  commentstyle=\color{purple}\ttfamily,
  stringstyle=\color{red}\ttfamily,
  sensitive=true
}

\lstdefinelanguage{Template}{
  keywords={include, import},
  morestring=[b]\|,
  morestring=[b]\|\|,
  showstringspaces=false,
  morecomment=[s]{'}{'},
  morecomment=[s]{"}{"},
  keywordstyle=\color{blue}\bfseries,
  ndkeywordstyle=\color{darkgray},
  identifierstyle=\color{black},
  commentstyle=\color{red}\ttfamily,
  stringstyle=\color{blue}\ttfamily\bfseries,
  sensitive=true
}

\title{DQM Tools and Techniques of the SND Detector}

\author[1,2]{K.V. Pugachev\note{Corresponding author.},}
\author[1,2]{T.V. Dimova,}
\author[1,2]{L.V. Kardapoltsev,}
\author[1,2]{A.A. Korol,}
\author[1]{D.P. Kovrizhin,}
\author[1]{and D.A. Shtol}

\affiliation[1]{Budker Institute of Nuclear Physics, SB RAS,\\ Novosibirsk, 630090, Russia}
\affiliation[2]{Novosibirsk State University,\\ Novosibirsk, 630090, Russia}

\emailAdd{K.V.Pugachev@inp.nsk.su}

\abstract{SND detector operates at the VEPP-2000 collider (BINP, Novosibirsk). To improve events selection for physical analysis and facilitate online
detector control we developed new data quality monitoring (DQM) system. 
The system includes online and reprocess control modules, automatic decision making scripts, interactive (web based) and program (python) access to various quality estimates. This access is implemented with node.js server with data in RDBMS MySQL. We describe here general system logics, its components and some implementation details.}


\proceeding{The International Conference ``Instrumentation for Colliding Beam Physics'' (INSTR20)\\
  24 -- 28 February, 2020\\
  Novosibirsk, Russia}

\begin{document}
\maketitle
\flushbottom

\section{Introduction}

The SND detector \cite{SNDAchasov,SNDAbramov,SNDAulchenko} operates at the VEPP-2000 collider \cite{VEPPKhazin} since 2008. It produces hundreds gigabytes of stored raw data per day~\cite{DAQupd14}. The data are complemented with dozens megabytes of metadata, facility conditions (beam energy, luminosity, crate temperatures, etc.) and additional statistics (histograms etc.) that could be used in reconstruction, processing and system control.

An important part of experiment software is a data quality monitoring (DQM) system. This is necessary to obtain the meaningful data for physical analysis and to control the detector state.  Recently DQM software was seriously redesigned. We present here the new system and its first usage experience.

The data quality metadata are generated at every stage of data collecting and reprocessing. The new DQM system includes software tools that
\begin{itemize}
\item show data acquisition summary and histograms;
\item collect quality data from automated scripts and users input;
\item have hierarchical quality model;
\item support several parameter sets for different stages of data processing;
\item provide quality information getter UI's and API's.
\end{itemize}

\section{Estimating Run Quality}

The minimal collection of events for analysis is referred to here and below as ``\emph{run}''. This is also the minimal unit for the data quality estimation. Parameters to monitor are defined by the detector subsystem experts with optional scripts which assign data quality marks to runs according to configuration. These marks could be ``bad'', ``user has to decide'', ``in doubt'', ``good'', ``no data''. They could be also assigned manually by dedicated persons which  are usually operators, run coordinator and detector subsystems experts.

\subsection{The Data Acquisition Stage}

At this stage an operator monitors the experiment data quality right after the data acquisition. A large set of histograms (e.g. drift chamber layers statistics, calorimeter energies distribution) becomes available minutes/hours later after the data are processed by a high level trigger and recorded. Our DQM system then launches predefined scripts which assign quality marks where possible, and then displays the histograms and automatic quality marks to an operator. The operator shall check them, assign the quality marks which were not set automatically, and, if necessary, correct automatic marks. Not all important quality parameters yet covered with automatic decision scripts. The DQM system requires that an operator fills the gaps. In order to help one to do it without special knowledge, the interface displays reference histograms. Subsystem experts also may leave comments about their decision like for example ``the histogram has to have two peaks''. Having checked all the parameters an operator could either proceed with other activities or report a problem to a run coordinator the same day it appeared.

During the data collection dedicated person (the run coordinator) makes sure that the data collection goes smoothly. This person keeps an eye on the operators checking quality data. The DQM system provides a day summary and a month view for this purpose in addition to the individual run view.

Interactive DQM interface is implemented as a web application. So the run coordinator and the experts can remotely discuss the quality and make sure the detector works fine.

A list of good runs could be exported also for prompt calibration programs to use.

Checks performed at this stage:
\begin{itemize}
\item Check the run validity: enough time, enough events, good collider currents etc.
\item Check the detector subsystems: calorimeter, tracking system, aerogel counters, muon system, trigger electronics etc.
\end{itemize}

\subsection{The Reprocessing Stage}

The second stage of the data quality control is done when data has been reprocessed for analysis. At that time we have more information, including that available only after completion the experiment. Reprocessing software applies proper and final calibration (conditions) data and produces new meta-statistics (histograms, counters, averages) to check. These meta-data are analysed later with related scripts based on configuration. 

Data preparation requires creative approach and immersion for several days or even weeks. The person who deals with this task have more general view than individual runs. It could be necessary to make quality decisions based on run ranges or run sets at once. Now \emph{the DQM system provides a single point of storing, discussing and retrieving quality information}. The data experts can easily access the first stage quality data or investigate some data mysteries with the detector subsystem experts. Having done that, we can export a list of good runs for processing.

Checks of the stage could include
\begin{itemize}
\item check the run validity
  \begin{itemize}
  \item enough run time, enough events,
  \item good event number ratio for $e^{+}e^{-} \to e^{+}e^{-}$, $e^{+}e^{-} \to \gamma\gamma$;
  \end{itemize}
\item check the subsystems, examine specific runs in detail.
\end{itemize}

\section{Interaction With Users}

An operator, a physicist or subsystem expert can interact with the DQM system using three interfaces: web interface, program getter access and DQM scripts respectively. In principle direct access to the DQM database is also possible, but this way is generally discouraged for any use other than system administration and development.

\subsection{Web Interface}

\begin{figure}[htbp]
\centering
\begin{minipage}{12pc}
\includegraphics[width=12pc]{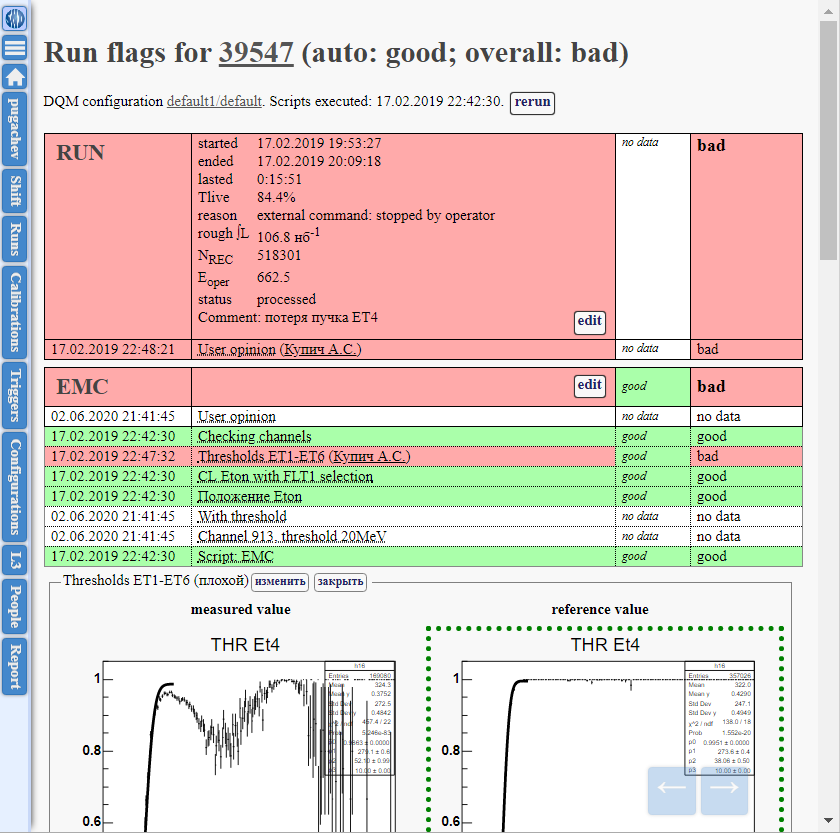}
\end{minipage}\quad\quad
\begin{minipage}{12pc}
\includegraphics[width=12pc,clip]{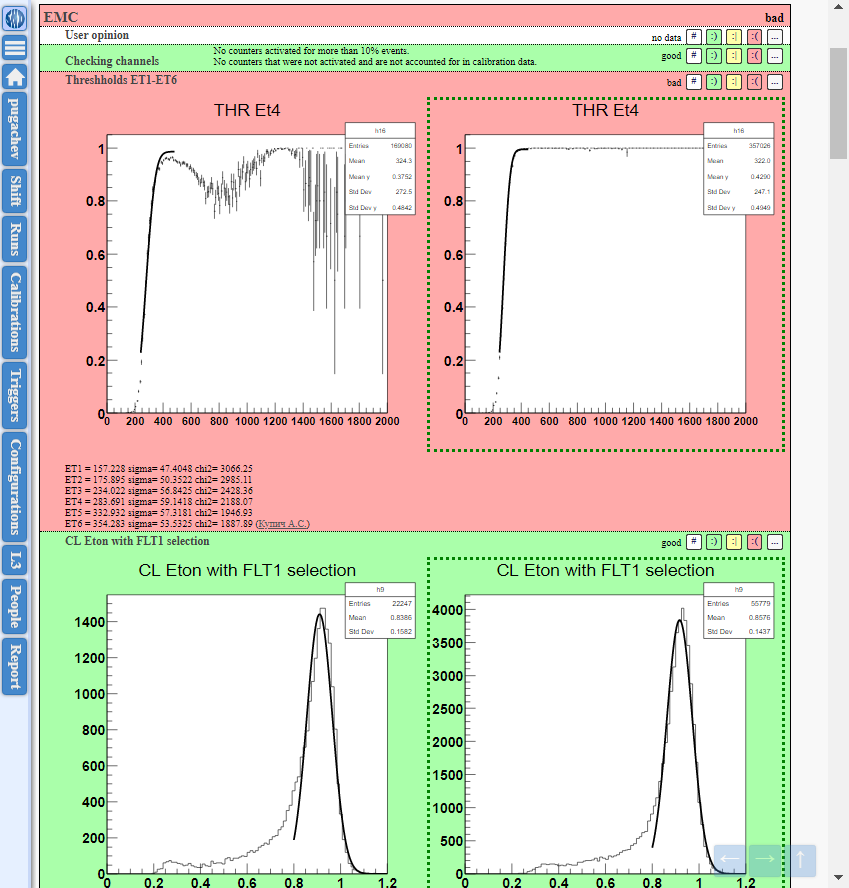}
\end{minipage} 
\caption{\label{PicDQMView}Reviewing a run quality -- the expert mode and the operator mode (translated in English for better reader experience).}
\end{figure}

This is the main way of manual interaction. It allows users to compare actual numbers and histograms with reference ones, to assign their quality marks and leave comments, to select runs by several quality criteria and to view or edit their quality data.
The interface provides different views for operators and experts.

The expert views are optimized for investigating quality data (figure~\ref{PicDQMView}, on the left). They contain forms for filtering runs by quality, run list view and particular run summary with parameters accessible by a mouse click.

The operator views  are optimized to check limited set of parameters for each run (figure~\ref{PicDQMView}, on the right).  It displays actual and reference representation (histograms, averages etc.) of pre-defined parameters set. An operator can monitor the run log and walk back and forth at the run quality view.

\subsection{Program Getter Access}

Data quality information is also available using program getter access from Python scripts. This language is chosen because embedded Python interpreter is used for configuration of experiment data processing framework~\cite{SUMO}. The interaction (figure~\ref{FigPythonGetter}) is relatively simple. The retrieved data (overall or per-system quality marks) could be used to filter qualified runs in automated calibration software or in analysis.

\begin{figure}[htbp]
  \begin{lstlisting}[language=python]
  # getting quality data for a single run 41000
  from RunQuality import DataQualityGetter
  getter = DataQualityGetter()
  quality = getter(41000)

  # getting quality data for multiple runs
  from RunQuality import DataQualityGetter
  getter = DataQualityGetter()
  quality, missed = getter.cache(range(41000, 41010))
  \end{lstlisting}
  \caption{\label{FigPythonGetter} Examples of using DQM python getter.}
\end{figure}

The program getter can either interact with web interface (figure~\ref{FigPythonGetter}, single run example), or load cached integral quality information (figure~\ref{FigPythonGetter}, multiple runs example) from the database. The single mode could be used for triggering quality estimation when login credentials are provided. The multiple mode is faster. However it may result in missing some information if some cache entries are expired or don't exist.

\subsection{DQM Scripts}

\begin{figure}[htbp]
  \begin{lstlisting}[language=C++]
  void script_example(
    int run, // this run number
    const char * hists // histograms ROOT file path or NULL
  ) {
    if(hists == NULL) {
      parameter("param1")
        .quality(QBAD)
        .comment("No histograms!")
        .valueNull().refNull(); // unset values
    } else {
      // check the histograms somehow (e.g. by rolling dice)
      parameter("param1")
        .quality(gRandom->Integer(2) ? QGOOD : QBAD
        .valueHist("CL/h29")
        .comment("Checked using lazy Monte-Carlo method.");
    }
    
    // apply the changes, set script execution status
    flush_parameters(QGOOD);
  }
  \end{lstlisting}
  \caption{\label{FigDQMScript} A simple DQM script.}
\end{figure}

Automated data quality estimation is performed by executing special scripts. A script (like at figure~\ref{FigDQMScript}) is a ROOT~\cite{ROOT} macro that accepts several parameters like run number, histogram, file path, etc. The script shall analyze the data and assign quality marks to related subsystem for one run. It also can set parameters titles/comments, choose custom histograms/numbers to show or even hide them. These actions are performed using a simple C++ API.

The system provides the infrastructure. Usually detector subsystem experts create their scripts based on their understanding.

These scripts are executed by a server when an authorized user accesses a web page containing run quality data. Having executed the scripts a user can review the results and correct some data if necessary at any time. The output produced by scripts is cached, so they are executed when accessing for the first time. Cache invalidation is available for experts.

\section{Implementation Details}

\begin{figure}[htbp]
\centering
\includegraphics[width=30pc]{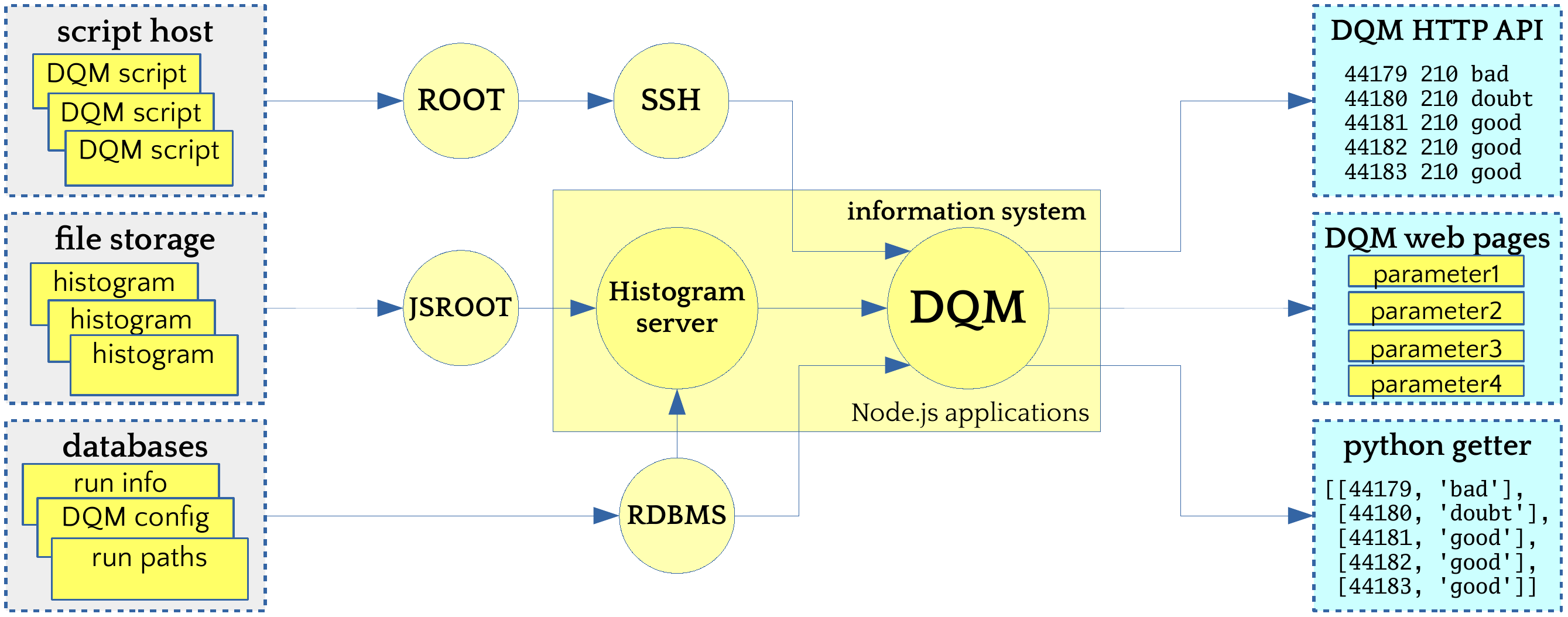}
\caption{\label{PicImplementation}The DQM data flow.}
\end{figure}

The described software is implemented as an SND information system~\cite{MSIS2017} component. The server part is integrated to the Node.js~\cite{NodeJS} application (in JavaScript). The system uses SND databases running under MySQL RDBMS and node-mysql for accessing them. Please refer to figure~\ref{PicImplementation} for more details.

The histograms mentioned before are served by another Node.js application that uses JSROOT~\cite{JSROOT} both at server and client sides for reading histogram files and rendering histograms. The server can use also old CERNLIB HBOOK files converting them by h2root utility.

\section{Applying to Experiment Data}

Having put the new software into production, we implemented DQM scripts for several subsystems (calorimeter, muon system, trigger electronics). The new interface was used at the data acquisition stage in 2019.

\begin{figure}[htbp]
\centering
\includegraphics[width=32pc,clip]{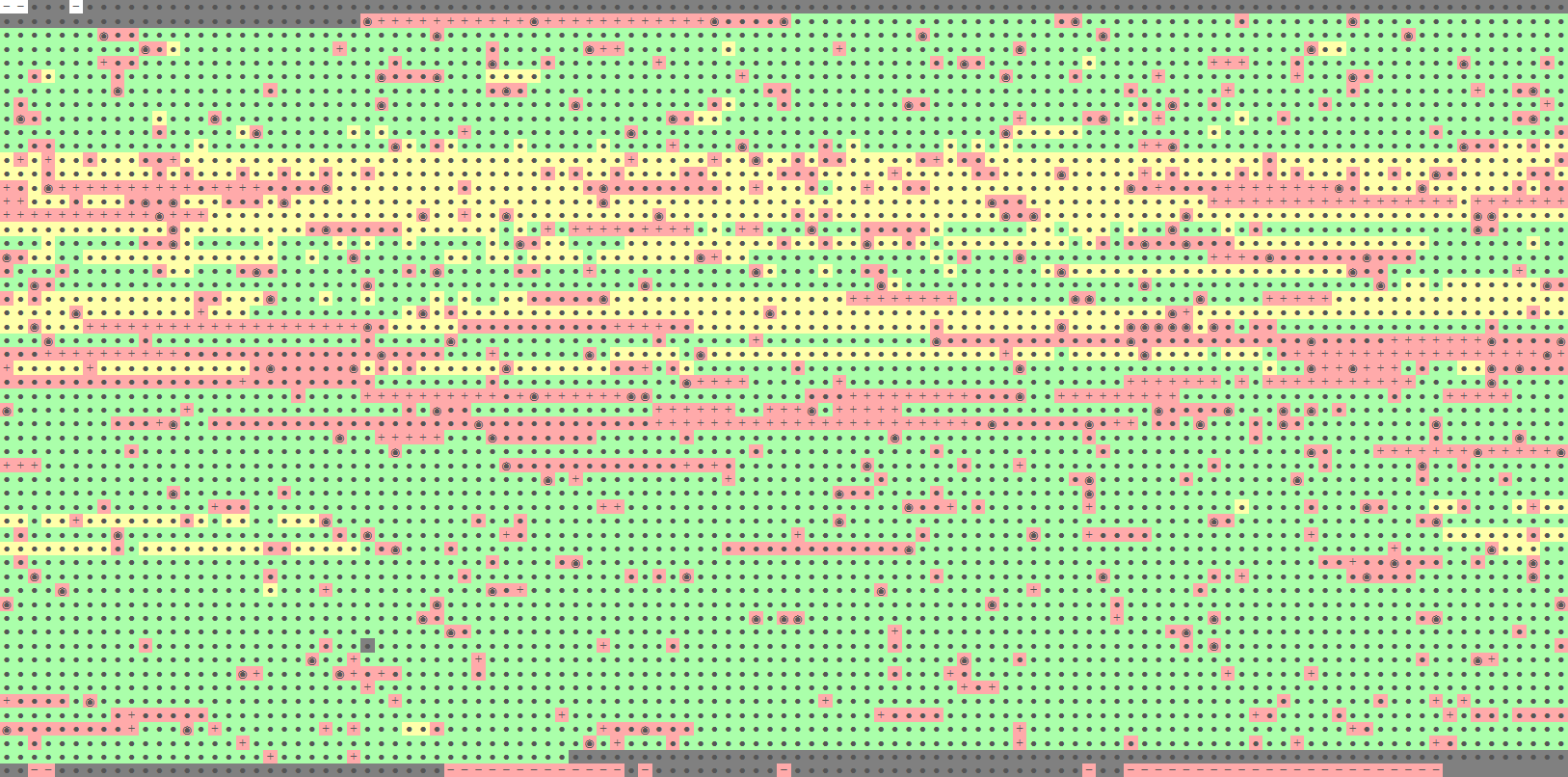}
\caption{\label{PicRHO2018}Integral quality data for RHO2018 experiment. Each cell represents a run (green, yellow and red backgrounds for good, tolerable and bad quality respectively).}
\end{figure}

At the same time, the data collected in the previous (2018) year were marked up during the reprocessing stage. More than five thousands runs (including cosmic ones) containing 3.1 billion stored events were checked. This check resulted in detecting 23\% runs having bad quality (cosmic, short, test or erroneous runs), 62\% good ones and 14\% ones with tolerable quality. The results are shown at the figure~\ref{PicRHO2018}. Please note that  bad runs tend to be significantly shorter in terms of time, events count and integral luminosity, however they occupy the same area on the figure.

\section{Conclusion}

A new version of the SND experiment DQM system provides a framework for automatic  and manual data quality estimation, interactive (web) and program user interfaces. SND data quality is monitored in data acquisition and processing stages with different goals and assumptions.

The new DQM system was put into production in 2019. The data acquisition configuration was used during data taking of 2019 and 2020. The reprocessing configuration was used for data of 2018 and 2019.


\acknowledgments

This work is partly supported by the RFBR grant 18-02-00382.


\bibliographystyle{JHEP}
\bibliography{INSTR20-pugachev}{}

\providecommand{\href}[2]{#2}\begingroup\raggedright\begin{thebibliography}{10}

\bibitem{SNDAchasov}
M.~N. Achasov et~al., \emph{{Spherical neutral detector for VEPP-2M collider}},
  \href{https://doi.org/10.1016/S0168-9002(99)01302-9}{\emph{Nucl. Instrum.
  Meth.} {\bfseries A449} (2000) 125}
  [\href{https://arxiv.org/abs/hep-ex/9909015}{{\ttfamily hep-ex/9909015}}].

\bibitem{SNDAbramov}
G.~N. Abramov et~al., \emph{{SND upgrade}}, {\emph{eConf} {\bfseries C010430}
  (2001) T10} [\href{https://arxiv.org/abs/hep-ex/0105093}{{\ttfamily
  hep-ex/0105093}}].

\bibitem{SNDAulchenko}
V.~M. Aulchenko et~al., \emph{{SND tracking system: Tests with cosmic muons}},
  \href{https://doi.org/10.1016/j.nima.2008.08.099}{\emph{Nucl. Instrum. Meth.}
  {\bfseries A598} (2009) 102}.

\bibitem{VEPPKhazin}
{\scshape CMD-3, SND} collaboration, \emph{{Detectors and physics at
  VEPP-2000}}, \href{https://doi.org/10.1016/j.nima.2010.02.246}{\emph{Nucl.
  Instrum. Meth.} {\bfseries A623} (2010) 353}.

\bibitem{DAQupd14}
A.~G. Bogdanchikov et~al., \emph{{SND data acquisition system upgrade}},  in
  \emph{{Proceedings, International Conference on Instrumentation for Colliding
  Beam Physics (INSTR14): Novosibirsk, Russia, February 24-March 1, 2014}},
  2014, \href{https://doi.org/10.1088/1748-0221/9/06/C06013}{DOI}
  [\href{https://arxiv.org/abs/1404.0490}{{\ttfamily 1404.0490}}].

\bibitem{SUMO}
D.~Bukin et~al., \emph{Snd off-line framework},  in \emph{CHEP 2001:
  Proceedings of the International Conference on Computing in High Energy and
  Nuclear Physics September 3-7, 2001 Beijing, P.R. China}, (New York),
  pp.~145--148, Science Press, 2001.

\bibitem{ROOT}
R.~Brun and F.~Rademakers, \emph{{ROOT: An object oriented data analysis
  framework}}, \href{https://doi.org/10.1016/S0168-9002(97)00048-X}{\emph{Nucl.
  Instrum. Meth.} {\bfseries A389} (1997) 81}.

\bibitem{MSIS2017}
K.~Pugachev and A.~Korol, \emph{{Management system for the SND experiments}},
  \href{https://doi.org/10.1088/1748-0221/12/09/C09006}{\emph{JINST} {\bfseries
  12} (2017) C09006} [\href{https://arxiv.org/abs/1705.04317}{{\ttfamily
  1705.04317}}].

\bibitem{NodeJS}
``{Node.js}.'' \url{https://nodejs.org/en/}.

\bibitem{JSROOT}
B.~Bellenot and S.~Linev, \emph{{JavaScript ROOT}},
  \href{https://doi.org/10.1088/1742-6596/664/6/062033}{\emph{J. Phys. Conf.
  Ser.} {\bfseries 664} (2015) 062033}.

\end{thebibliography}\endgroup
\end{document}